\documentstyle[emulateapj]{article}
\lefthead{Hibbard \& Yun}
\righthead{Luminosity Profiles of Merger Remnants}
\begin{document}

\title{Luminosity Profiles of Merger Remnants} 

\author{J.~E.~Hibbard}
\affil{National Radio Astronomy Observatory$^1$,
520 Edgemont Road, Charlottesville, VA 22903-2475 ({\it jhibbard@nrao.edu})}
\altaffiltext{1}{The National Radio Astronomy Observatory is a facility of 
the National Science Foundation operated under cooperative agreement 
by Associated Universities, Inc.}

\and 

\author{Min.~S.~Yun}
\affil{National Radio Astronomy Observatory$^1$,
	P.O. Box 0, Soccoro, NM~~87801-0387 ({\it myun@nrao.edu})}

\medskip

\begin{abstract}

Using published luminosity and molecular gas profiles of the
late-stage mergers NGC 3921, NGC 7252 and Arp 220, we examine the
expected luminosity profiles of the evolved merger remnants,
especially in light of the massive CO complexes that are observed in
their nuclei. For NGC 3921 and NGC 7252 we predict that the resulting
luminosity profiles will be characterized by an $r^{1/4}$ law. In view
of previous optical work on these systems, it seems likely that they
will evolve into normal ellipticals as regards their optical
properties.  Due to a much higher central molecular column density,
Arp 220 might not evolve such a ``seamless'' light profile. We
conclude that ultraluminous infrared mergers such as Arp 220 either
evolve into ellipticals with anomalous luminosity profiles, or do not
produce many low-mass stars out of their molecular gas complexes.

\end{abstract}


\section{Introduction}

The merger hypothesis for elliptical galaxy formation, as put forth by
Toomre \& Toomre (1972; see also \cite{Toomre77}), posits that two
spiral galaxies can fall together under their mutual gravitational
attraction, eventually evolving into an elliptical-like remnant.  An
early objection to this hypothesis was that the cores of ellipticals
are too dense to result from the dissipationless merging of two
spirals (\cite{Carlberg86,Gunn87,HSH93}).  An obvious solution to this
objection is to include a dissipative (gaseous) component in the
progenitors (for other solutions, see
e.g.~\cite{VW85,Barnes88,Lake89}).  Numerical experiments including
such a component readily showed that gaseous dissipation can
efficiently drive large amounts of material into the central regions
(\cite{Negroponte83,Noguchi86,BarnesH91}).  Indeed, it was considered
a great success for early hydrodynamical work to be able to reproduce
dense knots of gas within the central regions of simulated merger
remnants, similar to the gas concentrations observed in IR
luminous mergers (\cite{BarnesH91,San88b}).  Since the inferred
central gas mass densities in the observed gas knots are comparable to
the stellar mass density seen in the cores of normal ellipticals
($\sim 10^2 M_\odot$ pc$^{-3}$), it seems natural that they
could be the seed of a high surface brightness remnant 
(\cite{Kormendy92}).

Subsequent numerical work by Mihos \& Hernquist (1994; hereafter MH94)
finds that the dissipative response of the simulated gas component is
so efficient that the resulting mass profiles of the simulated
remnants are unlike those seen in normal ellipticals. In particular,
ensuing starformation leaves behind a dense stellar core whose surface
density profile does not join smoothly onto the de Vaucoleurs
$r^{1/4}$ profile of the pre-existing stellar population.  Instead,
the profiles exhibit a ``spike'' at small radii, with a suggested
increase in surface brightness by factors $\sim$100.  While the
predicted break in the mass density profile occurs at spatial scales
comparable to the gravitational softening length of the simulations,
making the precise slope somewhat questionable, the conclusion that
the profiles should exhibit a clear break was considered firm
(MH94). This prediction, if confirmed, offers a means to constrain the
frequency of highly dissipative mergers in the past by searching for
their fossil remnants in the cores of nearby ellipticals. However, the
numerical formalisms used in MH94 to model gaseous dissipation, star
formation, and energy injection back into the ISM from massive stars
and SNe (``feedback") are necessarily ad hoc in nature.  As such,
these predictions should be viewed as preliminary (as Mihos \&
Hernquist themselves note).

We investigate this question observationally by converting the
observed gas column densities of on-going or late-stage mergers into
optical surface brightness by assuming a stellar mass-to-light ratio
appropriate for an evolved population.  This light is added to the
observed luminosity profile after allowing it to age passively.  The
resulting luminosity profile is examined for anomalous features such
as the sharp break predicted by MH94. 

\begin{figure*}
\epsscale{1.0}
\plotone{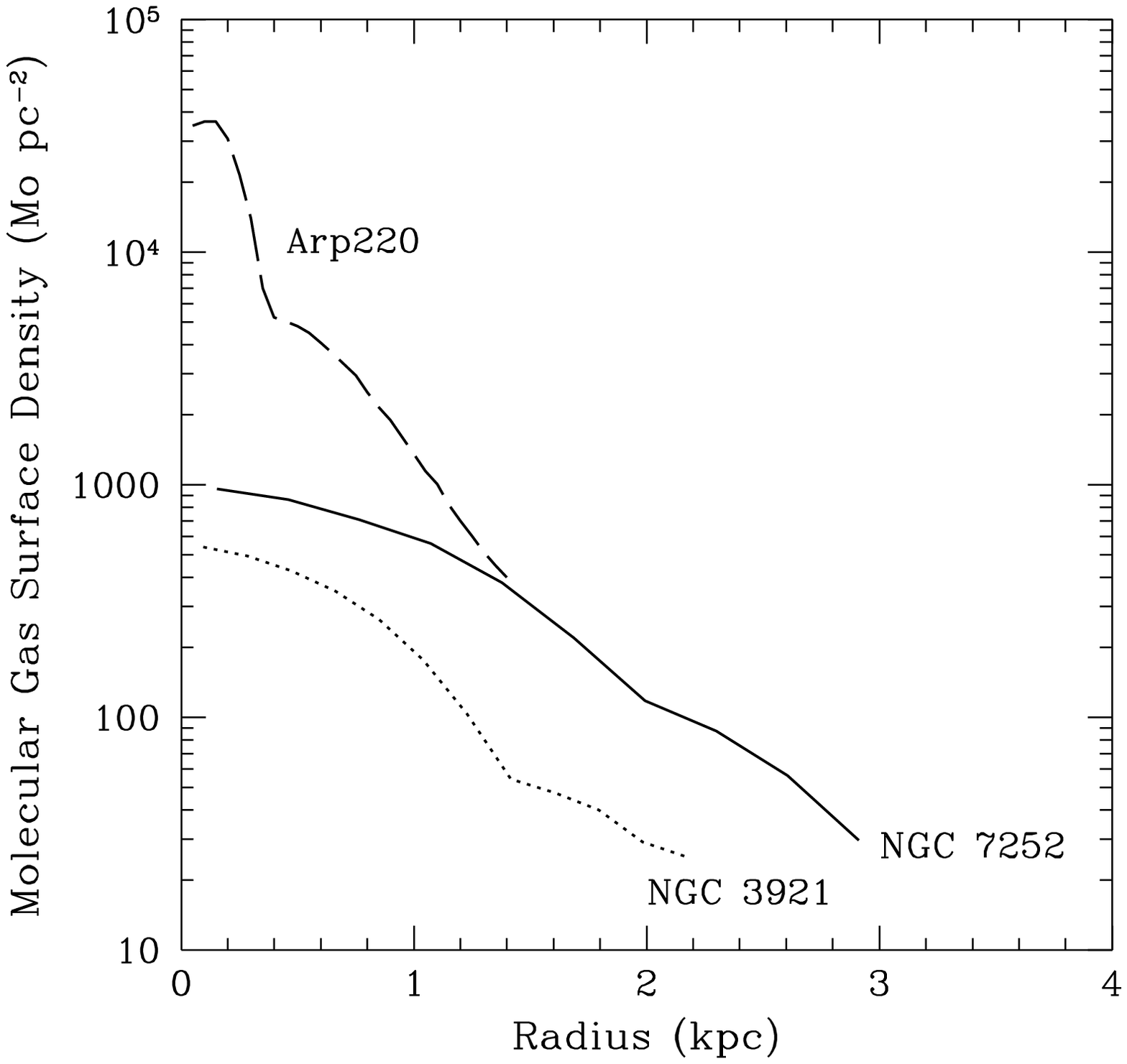}
\caption
{Radial surface molecular gas mass density profiles 
(in $M_\odot \, {\rm pc}^{-2}$) derived
from the integrated CO flux maps are plotted as a function
of radius. The CO fluxes have been turned into mass densities by
adopting a conversion factor of $N_{H_2} /I_{CO} = 3\times10^{20} \,
{\rm cm^{-2}\,(K \, km \, s^{-1})^{-1}}$. 
The data are from Yun \& Hibbard (1999a) for NGC 3921 (dotted line); 
from Wang et al.~(1991) for NGC 7252 (solid line); and 
from Scoville et al.~(1997) for Arp 220 (dashed line). 
\label{fig:radplotA}}
\end{figure*}

We conduct this experiment with the late-stage mergers NGC 3921 and
NGC 7252 and with the ultraluminous infrared (ULIR) merger Arp 220.
These systems were chosen because they have been observed in the
CO(1-0) molecular line transition with resolutions (full width at half
maximum) of $2^{\prime\prime}-2.5^{\prime\prime}$
(\cite{YH99a,Wang92,Sco97}).  The resulting spatial radial resolution
(300--400 pc; $H_o$= 75 km s$^{-1}$ Mpc$^{-1}$) is similar to the
hydrodynamical smoothing length used in MH94 ($\sim$ 350
pc)\footnote{Assuming scaling parameters appropriate for a
MilkyWay-like progenitor.}, indicating that the molecular line
observations have sufficient resolution to resolve the types of mass
concentrations found in the simulations.  The molecular gas surface
densities of each of these systems are plotted in
Figure~\ref{fig:radplotA}, converted from CO fluxes by adopting a
conversion factor of $N_{H_2} /I_{CO} = 3\times10^{20} \, {\rm
cm^{-2}\,(K \, km \, s^{-1})^{-1}}$ (\cite{Young91}).  Detailed
studies on each of these systems, which fully discuss their status as
late stage mergers, can be found in Schweizer (1996) and Hibbard \&
van Gorkom (1996) for NGC 3921; Schweizer (1982) and Hibbard et
al.~(1994) for NGC 7252; and Scoville et al.~(1997) 
for Arp 220.

\section{Results}

\subsection{NGC 3921 \& NGC 7252}

For these moderately evolved merger remnants (ages of $\sim$ 0.5--1
Gyr since their tidal tails were launched, \cite{Hib94,HvG96}), the
observed gas and luminosity profiles are used to predict the expected
luminosity profile of a 2 Gyr old remnant. We assume that all of the
molecular gas is turned into stars at the same radii, adopting an
exponentially declining starformation history. The present luminosity
profile is allowed to fade due to passive aging effects, and the final
luminosity profile is the sum of these two populations. The molecular
gas profiles plotted in Fig.~\ref{fig:radplotA} are converted into gas
mass densities by multiplying by a factor of 1.36 to take into
consideration the expected contribution of Helium. Optical luminosity
profiles have been obtained by Schweizer (1982, 1996), Whitmore et
al.~(1993), and Hibbard et al.~(1994), showing them to be well fitted
by an $r^{1/4}$ profile over all radii, with no apparent luminosity
spikes.

The gas surface densities ($\Sigma_{gas}$ in $M_\odot\,{\rm pc}^{-2}$)
are converted to optical surface brightnesses ($\mu_B$) by dividing by
the stellar mass-to-light ratio ($M_*/L_B$) expected for a 2 Gyr old
population. We adopt the stellar mass-to-light ratios given by de Jong
(1995, Table 1 of ch.~4), which were derived from the population
synthesis models of Bruzual \& Charlot (1993) for an exponentially
declining star formation history, a Salpeter IMF, and Solar metallicity
($M_*/L_B = 0.82 \, M_\odot L_\odot^{-1}$ at 2 Gyr). Noting that 1
$L_\odot \,{\rm pc}^{-2}$ corresponds to $\mu_B$ = 27.06 mag
arcsec$^{-2}$ (adopting $M_{B,\odot}=+5.48$), the conversion from gas
surface density to optical surface brightness is given by $\mu_B(r) =
27.06\,{\rm mag \, arcsec}^{-2} - 2.5\times log[\Sigma_{gas}(r) /
(M_*/L_B) ]$.

The luminosity profiles of the evolved remnants are estimated from the
observed $B$-band profiles (\cite{Schwe96a,Hib94}), allowing for a
fading of +1 mag arcsec$^{-2}$ in the $B$-band over the next 2 Gyr
(\cite{BC93,Schwe96a}), and adding in the expected contribution of the
population formed from the molecular gas, calculated as above. We
emphasize that this should favor the production of a luminous
post-merger population, since it assumes the that none of the
molecular gas is lost to stellar winds or SNe and the adopted IMF 
favors the production of many long-lived low-mass stars.

\begin{figure*}
\epsscale{0.7}
\plotone{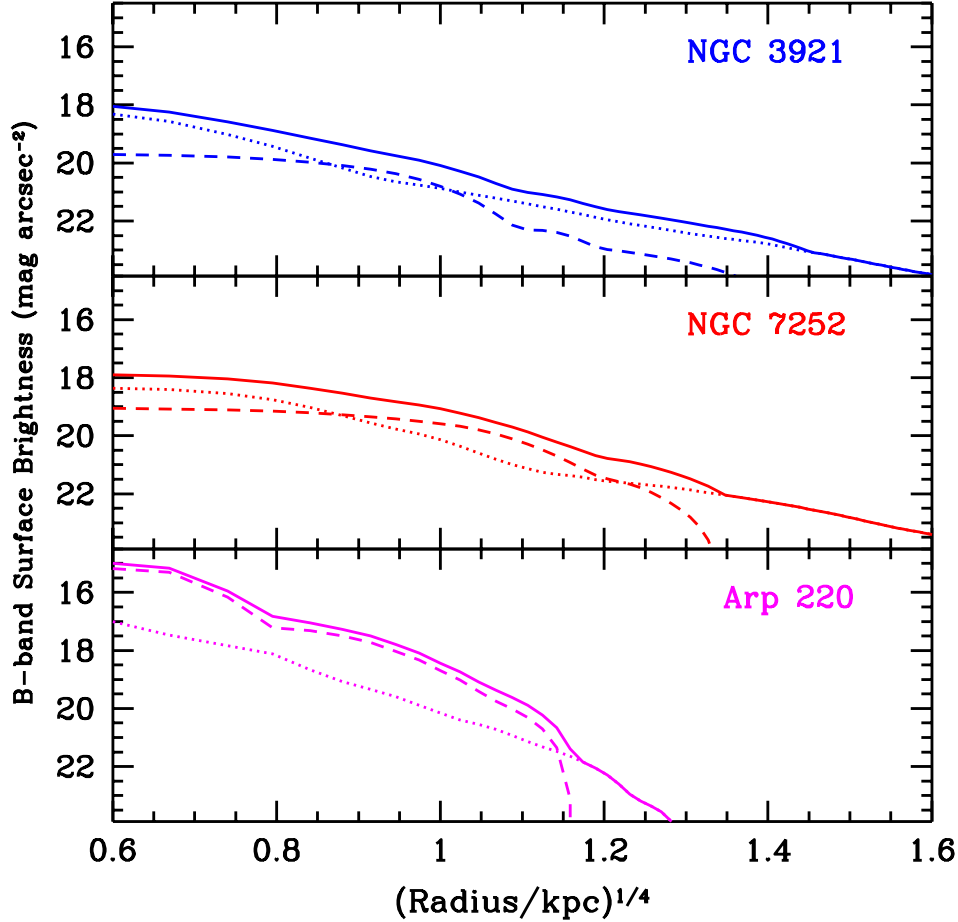}
\caption
{Predicted luminosity profiles as a function of the fourth-root of the
radius for the aged remnants of NGC 3921, NGC 7252, and Arp
220. Dotted lines: expected contribution to the $B-$band profile by
the present stellar component, after aging by 2 Gyr. For NGC 3921 and
NGC 7252 this is calculated from the presently observed $B-$band
luminosity profiles (taken from the data in Schweizer 1996 and Hibbard
et al.~1994) and fading by 1 mag arcsec$^{-2}$.  For Arp 220 we use
the observed $K-$band luminosity profile (from Scoville et al.~1998)
adopting $B-K=4$ for a 10 Gyr old population.  Dashed lines: expected
contribution to the $B-$band profile from the expected post-merger
population if all the molecular gas is turned into stars.  Solid
lines: the expected $B-$band luminosity profiles after adding in the
aged stellar and post-merger populations. No significant deviation
from a pure $r^{1/4}$ law expected for NGC 3921 or NGC 7252. Arp 220
does not show such a seamless profile, exhibiting a slight rise in at
small radii.
\label{fig:radplotB} }
\end{figure*}

The results of this exercise are plotted in Figure~\ref{fig:radplotB}.
This plot shows that the observed gas densities in NGC 3921 and NGC
7252, although high, are not high enough to significantly affect the
present luminosity profiles.  The profile of NGC 3921 is basically
indistinguishable from an $r^{1/4}$ profile. The profile of NGC 7252
does show a slight rise at small radii, but not the clear break
predicted by MH94. Therefore the resulting luminosity profiles of
these remnants are now and should remain fairly typical of normal
elliptical galaxies, and the conclusions of MH94 are not applicable to
{\it all} mergers of gas-rich galaxies.  Since both of these systems
also obey the Faber-Jackson relationship (\cite{Lake86}) and NGC 7252
falls upon the fundamental plane defined by normal ellipticals
(\cite{Hib94,Hib95}), we conclude that at least some mergers of
gas-rich systems can evolve into normal elliptical galaxies as far as
their optical properties are concerned.

\subsection{Arp 220}

Since Arp 220 is an extremely dusty object, its optical luminosity
profile is poorly suited for a similar analysis. Instead, we use a
luminosity profile measured in the near-infrared, where the dust
obscuration is an order of magnitude less severe. Arp 220 was recently
observed with camera 2 of NICMOS aboard the HST (Scoville et
al.~1998), and we use the resulting $K-$band luminosity profile,
kindly made available by N. Scoville. Since Arp 220 is presently
undergoing a massive starburst, the fading factor is much less certain
than for the already evolved systems treated above, and depends
sensitively on what fraction of the current light is contributed by
recently formed stars. We adopt a situation biased towards the
production of a discrepant luminosity profile by assuming that the
entire population was pre-existing, converting the observed $K-$band
profile to an evolved $B-$band profile by adopting a $B-K$ color of 4,
appropriate for a 10 Gyr old population (\cite{deJong}).  The
contribution due to the population formed from the molecular disk
is calculated exactly as before. The resulting profile is shown in
Figure~\ref{fig:radplotB}.

This figure shows that Arp 220 is predicted to evolve a luminosity
profile with a noticeable rise at small radii. This is due to the peak
in the molecular gas surface density at radii less than 0.5 kpc
(Fig.~\ref{fig:radplotA}). We conclude that Arp~220 has the potential
to evolve a similar feature in its luminosity profile, {\it if} indeed
all of the current molecular gas is converted into stars.  However,
the expected rise of $\sim$2 mag arcsec$^2$ in surface brightness (a
factor of $\sim$6) is considerably lower than the two orders of
magnitude increase predicted by the simulations (see Fig.~1 of MH94).

\section{Discussion}

From the above exercise, we conclude that neither NGC 3921 nor NGC
7252 are expected to show a significant deviation in their luminosity
profiles, and that the maximum rise expected for Arp 220 is
considerably lower than the two orders of magnitude increase predicted by
the simulations of MH94.  We conclude that the numerical formalisms
adopted in the simulations to treat the gas and star formation are
incomplete. Mihos \& Hernquist enumerate various possible shortcomings
of their code.  For example, their star-formation criterion is
extrapolated from studies of quiescent disk galaxies, and may not
apply to violent starbursts. Perhaps most importantly, their
simulations fail to reproduce the gas outflows seen in ULIR galaxies
(``superwinds" e.g.~\cite{HAM90,HLA93}), suggesting that the numerical
treatment of feedback is inadequate.

In spite of these results, it is still interesting that under some
conditions there might be an observational signature of a past merging
event in the light profile of the remnant.  The question is for which
mergers might this be the case?  Since $\Sigma_{H_2}$ is tightly
correlated with IR luminosity (\cite{YH99b}), we infer that only the
ultraluminous IR galaxies retain the possibility to evolve into
ellipticals with a central rise in their luminosity profiles. While
such profiles are not typical of ellipticals in general, they are not
unheard of. For example, $\sim$ 10\% of the the Nuker sample profiles
presented by Byun et al.~(1996) show such anomalous cores (e.g., NGC
1331, NGC 4239).  It is therefore possible that such systems evolve
from ultraluminous IR galaxies. This can be tested by careful
``galactic archaeology'' in such systems to search for signatures of a
past merger event (e.g., \cite{SS92,Malin97}).

However, it is not a foregone conclusion that systems like Arp 220 
will evolve anomalous profiles. This system presently hosts a very powerful
expanding ``superwind'' (\cite{Heckman96}), which may be able to eject
a significant fraction of the cold gas in a ``mass-loaded flow''
(e.g. \cite{Heckman99}). Such winds are common in ULIR galaxies
(\cite{HAM90,HLA93}). Another related possibility is that the IMF may
be biased towards massive stars (i.e.~``top heavy'',
\cite{Young86,Sco91b}). Such mass functions will leave far fewer stellar 
remnants than the IMF adopted here. A third possibility is that the
standard Galactic CO-to-H$_2$ conversion factor is inappropriate for
ULIR galaxies, and that the high gas surface densities derived from CO
observations (and thus the resulting stellar luminosity profile) may
be over-estimated (see \cite{Downes93,Bryant96}).

Some support for the idea that central gas cores may be depleted by
the starburst is given by a population synthesis model of NGC~7252,
which suggests that it experienced an IR luminous phase
(\cite{Uta94}).  While the current radial distribution of molecular
gas in NGC~7252 is flat and lacks the central core seen in Arp~220, it
appears to connect smoothly with that of Arp~220 in
Fig.~\ref{fig:radplotA}.  Therefore one may speculate that NGC~7252
did indeed have a radial gas density profile much like Arp~220 but has
since lost the high density gas core as a result of prodigious massive star
formation and/or superwind blowout.  However, the burst parameters are
not strongly constrained by the available observations, and a weaker
burst spread over a longer period may also be allowed
(\cite{Uta94}). Further insight into this question could be obtained
by constraining the past star formation history in other evolved
merger remnants.

In conclusion, a comparison of the peak molecular column densities and
optical surface brightnesses in NGC~3921 and NGC~7252 suggests that
some mergers between gas-rich disks {\it will} evolve into
elliptical-like remnants with typical luminosity profiles, even
considering their present central gas supply.  For ULIR galaxies like
Arp~220 the case is less clear. Such systems will either produce an
excess of light at small radii, as seen in a small number of
ellipticals, or require some process such as mass-loaded galactic
winds or a top-heavy IMF to deplete the central gas supply without
leaving too many evolved stars. If the latter possibility can be
excluded, then the frequency of such profiles may be used to constrain
the number of early type systems formed via ULIR mergers.\footnote{ We
note that any subsequent {\it dissipationless} merging of these cores
with other stellar system will tend to smooth out these profiles.}

\section{Summary}

\begin{itemize}

\item{Even under assumptions that favor the production of a 
luminous post-merger population, the dense molecular gas complexes
found in the centers of NGC 3291 and NGC 7252 should not significantly
alter their luminosity profiles, which are already typical of
elliptical galaxies (\cite{Schwe82,Schwe96a}). Since these systems
also obey the Faber-Jackson relationship (\cite{Lake86}) and NGC 7252
falls upon the fundamental plane defined by normal ellipticals
(\cite{Hib94,Hib95}), it appears that at least some mergers of
gas-rich systems can evolve into normal elliptical galaxies as far as
their optical properties are concerned.}

\item{The dense molecular gas complex found in the center of 
Arp 220 may result in a moderate rise in the remnants' luminosity
profile at small radii. Since the molecular gas column density is a
tight function of IR luminosity (\cite{YH99b}), we conclude that this
condition may apply to all of the ultraluminous infrared
galaxies. However, this does not preclude a merger origin for
elliptical galaxies since (1) About 10\% of the Nuker sample
ellipticals (Byun et al. 1996) show such rises in their radial light
profiles, and (2) it is possible that much of the gas in such systems
is blown into intergalactic space by the mass-loaded superwinds found
emanating from such objects (\cite{Heckman99,HLA93}).}

\item{The maximum expected rise in the luminosity profiles are 
considerably lower than the orders of magnitude increase predicted 
by the simulations (MH94). We therefore suggest that the 
numerical formalisms adopted in the simulations to treat the gas 
and star formation are incomplete.}

\end{itemize}

\acknowledgements

The authors thank N. Scoville for kindly providing the NICMOS K-band 
profile for Arp 220. We thank F. Schweizer and J. van Gorkom for 
comments on an earlier version of this paper, and R. Bender and 
C. Mihos for useful discussions, and the referee, J. Barnes, for a 
thorough report.

\clearpage

\end{document}